\documentclass[pre]{revtex4}
\usepackage[cp1251]{inputenc}
\usepackage[english,russian]{babel}

\usepackage{graphicx}
\usepackage{dcolumn}
\usepackage{eucal}
\usepackage[dvips]{epsfig}
\usepackage{amssymb}
\usepackage{amsmath}

\begin{document}


\newcommand{\bs}{\boldsymbol}
\newcommand{\mbb}{\mathbb}
\newcommand{\mcal}{\mathcal}
\newcommand{\mfr}{\mathfrak}
\newcommand{\mrm}{\mathrm}

\newcommand{\ovl}{\overline}
\newcommand{\p}{\partial}

\renewcommand{\d}{\mrm{d}}
\newcommand{\lap}{\triangle}

\newcommand{\lan}{\bigl\langle}
\newcommand{\ran}{\bigl\rangle}

\newcommand{\bse}{\begin{subequations}}
\newcommand{\ese}{\end{subequations}}

\newcommand{\be}{\begin{eqnarray}}
\newcommand{\ee}{\end{eqnarray}}

\newcommand{\ga}{\alpha}
\newcommand{\gb}{\beta}
\newcommand{\gc}{\gamma}
\newcommand{\gd}{\delta}
\newcommand{\gr}{\rho}
\newcommand{\eps}{\epsilon}
\newcommand{\veps}{\varepsilon}
\newcommand{\gs}{\sigma}
\newcommand{\gf}{\varphi}
\newcommand{\go}{\omega}
\newcommand{\gl}{\lambda}

\renewcommand{\l}{\left}
\renewcommand{\r}{\right}


\title{\bf TRUE DIELECTRIC AND IDEAL CONDUCTOR IN THEORY OF THE DIELECTRIC FUNCTION FOR COULOMB SYSTEM}
\author{V.B. Bobrov, S.A. Trigger}
\address{Joint\, Institute\, for\, High\, Temperatures, Russian\, Academy\,
of\, Sciences, 13/19, Izhorskaia Str., Moscow\, 125412, Russia;\\
email:\,satron@mail.ru}

\begin{abstract}
On the basis of the exact relations the general formula for the
static dielectric permittivity $\varepsilon(q,0)$ for Coulomb
system is found in the region of small wave vectors $q$. The
obtained formula describes the dielectric function
$\varepsilon(q,0)$ of the Coulomb system in both states in the
"metallic" \,state and in the "dielectric" \, one. The parameter
which determines possible states of the Coulomb system - from the
"true" \,dielectric till the "ideal" \,conductor is found. The exact relation for the pair correlation function
for two-component system of electrons and nuclei $g_{ec}(r)$ is found for
the arbitrary thermodynamic parameters.\\

PACS number(s): 64.10.+h, 05.70.Ce, 52.25.Kn, 64.60.Bd\\

\end{abstract}

\maketitle

1. \textbf{Long wavelength dielectric function for Coulomb system:
between dielectric and metallic states}.

We consider the static dielectric permittivity $\varepsilon(q,0)$
for the homogeneous and isotropic Coulomb system. The function
$\varepsilon(q,0)$ is determined as the proportionality
coefficient between the potential $U^{tot}(q,0)$ of the total
electric field in the medium and the external field potential
$U^{ext}(q,0)$ [1]
\begin{eqnarray}
U^{tot}(q,0)=\frac{U^{ext}(q,0)}{\varepsilon(q,0)}. \label{A1}
\end{eqnarray}

According to [2,3] the function $\varepsilon(q,0)$ is connected with the static polarization operator $\Pi(q,0)$,
which determines the response of the system on the screened external field by the relation
\begin{eqnarray}
\varepsilon(q,0)=1-\frac{4\pi}{q^2}\Pi(q,0), \label{A2}
\end{eqnarray}
\begin{eqnarray}
\Pi(q,0)=\sum_{a,b}z_a z_b e^2\Pi_{a\,b}(q,0), \label{A3}
\end{eqnarray}
where $z_a e$, $m_a$ and $n_a$ are the charge, the mass and the
average density of the particles of the sort $a$ in the system
with the chemical potentials $\mu_a$ at temperature $T$. The
system is considered under the condition of quasineutrality
\begin{eqnarray}
\sum_{a} e z_a n_a=0 \label{A4}
\end{eqnarray}

The functions $\Pi_{ab}(q,0)$ are the partial polarization
operators of the particle species $a$ and $b$. In diagram
technique [2,3] the functions $\Pi_{ab}(q,0)$  are the irreducible
(on one-line of the Coulomb interaction in the q-channel) parts of
the appropriate "density-density"\, Green functions
$\chi_{ab}(q,0)$, which determine response of the system on an
external field. In contrast with the Green functions
$\chi_{ab}(q,0)$ of the Coulomb systems, the polarization
functions $\Pi_{ab}(q,0)$ do not consist the singularities and are
the smooth functions in the region of small wave vectors $q$ (at
least for the normal systems). Therefore, the functions
$\Pi_{ab}(q,0)$ for the small values of $q$ can be represented in
the form
\begin{eqnarray}
\Pi_{a\,b} (q,0)\simeq \pi_{a\,b}^{(0)}+q^2 \pi_{a\,b}^{(2)},
\label{A5}
\end{eqnarray}

\begin{eqnarray}
\pi_{a\,b}^{(0)}=lim_{q\rightarrow 0} \Pi_{a\,b} (q,0), \;
\pi_{a\,b}^{(2)}=lim_{q\rightarrow 0} \left[\frac{\Pi_{a\,b}
(q,0)- lim_{q\rightarrow 0} \Pi_{a\,b} (q,0)}{q^2}\right].
\label{A6}
\end{eqnarray}

In [4,5] it was  shown that
\begin{eqnarray}
\pi_{a\,b}^{(0)}=-\left(\frac{\partial n_a}{\partial \mu_b}\right)_T. \label{A7}
\end{eqnarray}

Eq.~(\ref{A7}) is the generalization for the many-component
Coulomb system the well known result for the model case of the
one-component electron liquid, where this kind of equality is
called "the sum rule for compressibility" [6]
\begin{eqnarray}
\pi_{e\,e}^{(0)}=-\left(\frac{\partial n_e}{\partial
\mu_e}\right)_T=-n_e^2 K_T^e, \,\,
K_T=-\frac{1}{V}\left(\frac{\partial V}{\partial P}\right)_T,
\label{A8}
\end{eqnarray}
where V,  P and $K_T$ are respectively the volume, pressure and
the isothermal compressibility of the Coulomb system. In one's
turn, for the two-component Coulomb system which consists the
electrons (index - $e$) and nuclei (index - $c$), the limiting
relations [4,5] for the static structure factors $S_{a\,b} (q)$
can be found on the basis of  Eq.~(\ref{A8})
\begin{eqnarray}
lim \,S_{c\,c}(q\rightarrow 0)=n_c T K_T; \;\; lim \,S_{c\,c}(q\rightarrow 0)=\frac{n_c}{n_e}\,lim\, S_{e\,e}(q\rightarrow 0)=
\left(\frac{n_c}{n_e}\right)^{1/2}lim \,S_{e\,c}(q\rightarrow 0) \label{A9}
\end{eqnarray}

The functions $S_{a\,b} (q)$ are measured directly, including the
critical point region [7], in the experiments on the neutron
scattering. By inserting (\ref{A5}), (\ref{A6}) in  (\ref{A2}) and
(\ref{A3}) we obtain in the long-wavelength limit (the small
values of $q$) for the dielectric function of the Coulomb system
of an arbitrary composition the following relations
\begin{eqnarray}
\varepsilon(q,0)=\varepsilon_0^{st}+ \frac{\kappa^2}{q^2}, \;\;\; \kappa^2=-4\pi \sum_{a\,b}e^2 z_a z_b \pi_{a\,b}^{(0)}=
4\pi \sum_{a\,b}e^2 z_a z_b \left(\frac{\partial n_a}{\partial \mu_b}\right)_T  \label{A10}
\end{eqnarray}
\begin{eqnarray}
\varepsilon_0^{st}=1+4\pi \alpha, \; \;\; \alpha= -\sum_{a\,b}e^2 z_a z_b \pi_{a\,b}^{(2)}\label{A11}
\end{eqnarray}

It is evident, that all the coefficients in Eqs.~(\ref{A10}),
(\ref{A11}) are the functions of the thermodynamic parameters of
the Coulomb system. By use of the grand canonical ensemble one
easily arrive [8] at the equality
\begin{eqnarray}
T \left(\frac{\partial n_a}{\partial \mu_b}\right)_T =\frac{1}{V} <\delta N_a \delta N_b>, \;\;\; \delta N_a =N_a - <N_a>, \label{A12}
\end{eqnarray}
where $n_a=<N_a>/V$, $N_a$ is the operator of the total number of particles of the sort $a$ and the brackets $<...>$ means
the averaging on the grand canonical ensemble.

Inserting (\ref{A12}) in (\ref{A10}) and taking into account the quasineutrality condition we find for the Coulomb system at
the arbitrary parameters
\begin{eqnarray}
\kappa^2=\frac{4\pi}{T}\,\frac{<Z^2>}{V}\geq 0, \; \;\; Z=\sum_{a} z_a e N_a. \label{A13}
\end{eqnarray}
We have mention that the sign of the value $\alpha$ which is
introduced by Eq.~(\ref{A11}) is not determined in the moment. It
is easy to see from (\ref{A11}) that the value $\kappa$ coincides
in the appropriate limiting cases with the Debye and the
Thomas-Fermi wave vectors (see, e.g., [3]).

As an illustration, let us consider the action of the point charge
on the infinite homogeneous Coulomb system. Then, taking into
account (\ref{A1}), in $r$-space we obtain
\begin{eqnarray}
\frac{U^{tot}(r)}{U^{ext}(r)}= \frac{2}{\pi}\int_0^\infty \frac{d
q}{q}\,\frac{sin(q\,r)}{\varepsilon (q,0)} \label{A14}
\end{eqnarray}
In the limit $r\rightarrow \infty$ from Eqs.~(\ref{A10}),
(\ref{A14}) directly follows
\begin{eqnarray}
\frac{U^{tot}(r)}{U^{ext}(r)}\rightarrow\frac{1}{\varepsilon_0^{st}}\exp(-r/R_{scr}),\;\;\;
R_{scr}=\left(\frac{\varepsilon _0^{st}}{\kappa^2}\right)^{1/2},
\label{A15}
\end{eqnarray}
where $R_{scr}$ is the electrostatic field penetration length in
matter, or the screening radius, according to the terminology,
accepted in the theory of non-ideal plasma (see, e.g., [9]). This
value, as it follows from (\ref{A10}), characterizes the depth of
penetration for the electromagnetic field in the medium.

In the limiting case
\begin{eqnarray}
\frac{4\pi}{TV}\,<Z^2>\rightarrow 0, \,\,\, \kappa^2 \rightarrow 0
\label{A16}
\end{eqnarray}
the screening radius $R_{scr}$ tends to infinity
\begin{eqnarray}
R_{scr}\rightarrow \infty \label{A17}
\end{eqnarray}

Therefore, when the condition (\ref{A16}) is fulfilled, the
Coulomb system manifests itself as an "true" dielectric, which
changes only the amplitude of the electrostatic field on the value
$\varepsilon_0^{st}$ (\ref{A11}). In this sense the value
$\varepsilon_0^{st}$ can be treated as the dielectric constant of
the medium. Accordingly, the value $\alpha$ in (\ref{A11}) can be
considered as the electric polarization of the medium.

It is necessary to stress the essential circumstance. As in the
case of "traditional"\, consideration of the "metal-dielectric"\,
transition on basis of the analysis of the electron conductivity
(see, e.g., [10],[11]), one can maintain the relative character of
the division of matter on dielectrics and conductors, since all
dielectrics possess the non-zero conductivity for $T\neq 0$. The
similar statement is applied to the depth of penetration $R_{scr}$
(\ref{A15}) of the electrostatic field in matter. In metals the
penetration depth is very small in contrast with dielectrics,
where it can be of one order with the size of the system.

The indirect confirmation of this statement contains in [3], where
the generalized random phase approximation for calculation of the
polarization function $\Pi (q,0)$ is developed. This approximation
permits to take into account the bounded states of the electrons
and nuclei. In fact, it means the possibility to separate the
states in the Coulomb systems on "localized"\, and "delocalized"\
ones. In the last case the charged particles can spread on the
whole volume of the considering system.

In the opposite to (\ref{A16}) limiting case
\begin{eqnarray}
\frac{4\pi}{TV}\,<Z^2>\rightarrow \infty, \,\,\, \kappa^2
\rightarrow \infty,  \label{A18}
\end{eqnarray}
\begin{eqnarray}
R_{scr}\rightarrow 0. \label{A19}
\end{eqnarray}
If in the respective thermodynamic state the penetration length
$R_{scr}$ for the electrostatic field tends to zero the system can
be treated not simply as in the "metallic"\, state but as an
"ideal"\,conductor. In this limiting case the electrostatic field
cannot penetrate in the matter at all (in the limiting
interpretation, naturally).

Meanwhile the question arises on the relation between the true
"dielectric" and "ideal" conductor from one side and the static
conductivity $\sigma_{st}$ for the Coulomb system. In this
connection we have notice that according to the perturbation
theory of the diagram technique for Coulomb systems [2,3] the
charged particles interact by the the screening Coulomb potential
$U_{a\,b}^{scr}$
\begin{eqnarray}
U_{a\,b}^{scr}(q)=\frac{4\pi z_a z_b e^2}{q^2 \varepsilon(q,0)},
\label{A20}
\end{eqnarray}
which is similar to (\ref{A1}).

In the case when the limiting conditions (\ref{A16}) are fulfilled
the interaction potential $U_{a\,b}^{scr}(q)$ for small $q$ (or
for large distances) is similar to the initial Coulomb potential.
Therefore, one can suggest that the charged particles form the
collective "localized"\, state with the conductivity $\sigma_{st}$
equals to zero.

In the opposite limiting case (\ref{A18}) the initial Coulomb
potential is suppressed and one can suggest that the charged
particles are in the fully "delocalized"\, state with the
conductivity $\sigma_{st}$ tends to infinity.

Therefore, we can assert, that the representation of the
dielectric permittivity $\varepsilon (q,0)$ in the region of small
wave vectors $q$ in the form Eq.~(\ref{A10}) is universal and can
be used for description of the Coulomb system in both, the
"metallic"\, and the "dielectric"\, states of matter. The
parameter $<Z^2>/V$, changing from $0$ to $\infty$, determines the
the variety of the states of the Coulomb system - from the state
of the "true"\, dielectric and till the state of the "ideal"\,
conductor.

2. The analysis performed above is referred to the Coulomb systems
with two or more components of charged particles and to strong
inter-particle interaction. Since theoretical description of these
systems is difficult, the exact relations for the correlation
functions of such systems are very important and useful.

According to Eqs.~(\ref{A2}),(\ref{A3}),(\ref{A10}),(\ref{A13})
for the small values of wave vectors $q$ for the static dielectric
function $\varepsilon (q,0)$ the inequality
\begin{eqnarray}
\varepsilon(q,0)>1, \label{A21}
\end{eqnarray}
is fulfilled.

As shown in [12-14] from the inequality (\ref{A21}) follows , that
the dielectric permittivity $\varepsilon (q,\omega)$ satisfies to
the Kramers-Kronig relations
\begin{eqnarray}
Re \varepsilon(q,\omega)=1+P \int_{-\infty}^\infty \frac{Im
\varepsilon(q,\xi)}{\pi}\frac{d \xi}{\xi-\omega}, \label{A22}
\end{eqnarray}
\begin{eqnarray}
Im \varepsilon(q,\omega)=-P \int_{-\infty}^\infty \frac{Re
\varepsilon(q,\xi)-1}{\pi}\frac{d \xi}{\xi-\omega}. \label{A23}
\end{eqnarray}
Symbol $P$ means that  we consider the main value of the integral.
From (\ref{A22}) taking into account the relations [6]:
\begin{eqnarray}
Re \varepsilon(q,\omega)=Re \varepsilon(q,-\omega),\;\; Im
\varepsilon(q,\omega)=-Im \varepsilon(q,-\omega) \label{A24}
\end{eqnarray}
\begin{eqnarray}
Im \varepsilon(q,\omega)>0 \;\;\; \mbox{for}\;\; \omega>0
\label{A25}
\end{eqnarray}
is easy to see [15] that for the moments $m_n(q)$ of the
high-frequency ($\omega\rightarrow\infty$) expansion of the
function $Re \varepsilon(q,\omega)$
\begin{eqnarray}
Re \varepsilon(q,\omega)=1- \sum_{n=1}^\infty
\frac{m_n(q)}{\omega^{2n}}, \label{A26}
\end{eqnarray}
is fulfilled the condition
\begin{eqnarray}
m_n(q)= \frac{2}{\pi}\int_0^\infty \xi^{2n} Im \varepsilon(q,\xi)
d\xi. \label{A27}
\end{eqnarray}

According to the given above consideration the inequality
(\ref{A27}) is true in the long wavelength limit ($q \rightarrow
0$) for an arbitrary thermodynamic parameters in homogeneous and
isotropic Coulomb system.

Let us use for further consideration the expressions for two first
momenta $m_1(q)$ and $m_2(q)$ [16,17]
\begin{eqnarray}
m_1(q)=\omega_p^2, \label{A28}
\end{eqnarray}
\begin{eqnarray}
m_2(q)=\sum_a \omega_a^2 \left\{\frac{2T_a q^2}{m_a}+\frac{\hbar^2
q^4}{4m_a^2} \right\}+ \sum_{a\,,b}(n_a n_b)^{1/2} \int^\infty_0
k^2 (S_{a\,b}(k)-\delta_{a,\,b})\nonumber\\ \times \left\{
\frac{z_a^2 z_b^2 e^4}{m_a m_b} \left[\frac{(q^2-k^2)^2}{k q^3}
\ln \mid \frac{q+k}{q-k}\mid -\frac{2k^2}{q^2}+6 \right]-\frac{8
z_a^3 z_b e^4}{3m_a^2}\right\} d k, \label{A29}
\end{eqnarray}
Here $T_a$ is the exact average kinetic energy referred to one
particle of the sort $a$, $S_{a,\,b}(q)$ - the static structure
factor for the particles of the species $a$ and $b$, which is
connected with the pair correlation function $g_{a,\,b}(r)$,
\begin{eqnarray}
S_{a\,b}(q)=\delta_{a,b}+ (n_a n_b)^{1/2} \int \exp (i{\bf q
r})\{g_{a\,b}(r)-1\} d {\bf r}  \label{A30}
\end{eqnarray}
$\omega_p$ is the plasma frequency of the Coulomb system,
$\omega_a$ is the plasma frequency of the charges of the sort $a$
\begin{eqnarray}
\omega_a=\left(\frac{4\pi z_a^2e^2 n_a}{m_a}\right)^{1/2},
\,\,\omega_p= \left(\sum_a \omega_a^2 \right)^{1/2}. \label{A31}
\end{eqnarray}

For derivation of the relation (\ref{A29}) we used the potentials
of the Coulomb inter-particle interaction
\begin{eqnarray}
u_{a\,b}(q)=\frac{4\pi z_a z_b e^2}{q^2}. \label{A32}
\end{eqnarray}

From (\ref{A29}) follows [15]
\begin{eqnarray}
m_2(0)=\lim_{q\rightarrow 0} m_2(q)=\frac{8}{3} \sum_{a\,b}(n_a
n_b)^{1/2}\left\{\frac{z_a^2 z_b^2 e^4}{m_a m_b}-\frac{z_a^3 z_b
e^4}{m_a^2}\right\} \int^\infty_0 k^2 S_{a\,b}(k) dk \label{A33}
\end{eqnarray}
In the particular case of the two-component Coulomb system,
consisting the electrons (index $e$) and nuclei (index $c$), from
(\ref{A33}) under the quasi neutrality condition (\ref{A4}) we
arrive at the expression
\begin{eqnarray}
m_2(0)=\frac{\omega^4_e}{3}\left(1+\frac{z_c m_e}{m_c}\right)
\{g_{e\,c}(0)-1\} \label{A34}
\end{eqnarray}
Therefore, according to (\ref{A27}) in the homogeneous and
isotropic two-component Coulomb system the inequality
\begin{eqnarray}
g_{e\,c}(0)\geq 1 \label{A35}
\end{eqnarray}
has to be fulfilled for the arbitrary thermodynamic parameters.

The authors are thankful to the Netherlands Organization for
Scientific Research (NWO) for support of this work in the
framework of the grant № 047.017.2006.007.

\end{document}